\documentclass[preprint,showpacs,preprintnumbers,amsmath,amssymb]{revtex4-1}
\usepackage{mathrsfs}
\usepackage{graphicx}
\usepackage{dcolumn}
\usepackage{bm}

\begin{document}

\title{A key role of transversality condition in quantization of
photon orbital angular momentum}

\author{Chun-Fang Li\footnote{Email address: cfli@shu.edu.cn}}

\affiliation{Department of Physics, Shanghai University, 99 Shangda Road, 200444
Shanghai, China}

\affiliation{Shanghai Key Lab of Modern Optical System, Engineering Research Center of Optical Instrument and System, Ministry of Education, School of Optical-Electrical and Computer Engineering, University of Shanghai for Science and Technology, 200093 Shanghai, China}

\date{\today}

\begin{abstract}

The effect of the transversality condition on the quantization of the photon orbital angular momentum is studied. The quantum gauge that is deduced from the transversality condition is shown to be a Berry gauge. It determines a helicity-dependent reference point relative to which the position is canonically conjugate to the momentum. As a result, the photon orbital angular momentum about the origin of the laboratory reference system split into two parts. One is the orbital angular momentum of the photon about the reference point, which is canonical. The other is the orbital angular momentum of the photon concentrated at the reference point, which is dependent on the helicity. Only when the Berry-gauge degree of freedom of a paraxial beam is perpendicular to the propagation direction, does the total orbital angular momentum reduce to its canonical part. One of the observable effects of the Berry-gauge degree of freedom is also clarified.

\end{abstract}

\maketitle


\section{Introduction}

In their seminal work \cite{Allen} Allen and his co-researchers introduced the notion of eigenstate of the optical orbital angular momentum (OAM), which is described by a phase factor $\exp(i l \phi)$ with $l$ the eigenvalue. Since then a continuously growing number of theoretical and experimental works \cite{Arnaut, Mair, Picc, Franke, Leach, Vaziri, Leach04, Calvo, Nagali, Dada, Hies} appeared in the literature dealing with the eigenstate of the OAM at the level of single photons and its potential applications.
It is well known \cite{Sakurai} that the canonical quantization of the OAM $\hat{\mathbf L}$ is based on the canonical commutation relation,
\begin{equation}\label{CCR-L}
    [\hat{L}_i, \hat{L}_j] =i \hbar \epsilon_{ijk} \hat{L}_k,
\end{equation}
where Einstein's summation convention is assumed. This commutation relation in turn requires that the position $\hat{\mathbf X}$ and momentum $\hat{\mathbf P}$ satisfy the canonical commutation relations,
\begin{equation}\label{FQC}
[\hat{X}_i,\hat{X}_j] =0, \quad
[\hat{P}_i,\hat{P}_j] =0, \quad
[\hat{X}_i,\hat{P}_j] =i \hbar \delta_{ij}.
\end{equation}
That is to say, the position needs to be commutative.
Quantities $\hat{\mathbf X}$ and $\hat{\mathbf P}$ satisfying (\ref{FQC}) are known as the canonical position and canonical momentum \cite{Dirac}, respectively. They are called canonically conjugate \cite{Merz}. However, being extremely relativistic, the photon is nonlocal in position space \cite{Jauch-P, Amrein, Pauli, Rosewarne-S}.
Its position cannot be commutative \cite{Pryce, Newton, Wightman, Jordan, Hawton, Cohen}.
As a result, the OAM of the photon cannot be canonically quantized in accordance with the commutation relation (\ref{CCR-L}). So how do we understand the above mentioned eigenvalue of the OAM?

Due to its nonlocality, the position-space wavefunction for the photon cannot be defined \cite{Akhiezer, Bialynicki1996} in the usual sense \cite{Sakurai}. Only the momentum-space wavefunction can.
In a formalism that is not manifestly relativistic \cite{Cohen}, it is a vector function, $\mathbf{f}(\mathbf{k}, t)$, satisfying the Schr\"{o}dinger equation
\begin{equation}\label{SE-V}
    i \frac{\partial \mathbf{f}}{\partial t}= \omega \mathbf{f},
\end{equation}
where $\mathbf k$ is the wavevector, $\omega =ck$ is the angular frequency playing the role of the Hamiltonian, and $k=|\mathbf{k}|$.
Here there is one peculiar feature that does not usually occur in quantum mechanics. This is that the vector wavefunction is constrained by the transversality condition
\begin{equation}\label{TC}
    \mathbf{f} \cdot \mathbf{w}=0,
\end{equation}
where $\mathbf{w}=\mathbf{k}/k$ is the unit wavevector.
Because the Schr\"{o}dinger equation (\ref{SE-V}) together with the constraint (\ref{TC}) is equivalent to the relativistic Maxwell's equations \cite{Akhiezer, Cohen}, the constraint (\ref{TC}) can be viewed in quantum mechanics as expressing the relativistic nature of the photon.
In the vector representation, the operator for the OAM about the origin of a laboratory reference system is given by \cite{Akhiezer, Cohen, Li09-1, Li16}
\begin{equation}\label{L}
    \hat{\mathbf L}= -\hat{\mathbf P} \times \hat{\mathbf X},
\end{equation}
where
\begin{subequations}\label{XandP}
\begin{align}
  \hat{\mathbf P} &= \hbar \mathbf{k},    \label{P}\\
  \hat{\mathbf X} &= i \nabla_\mathbf{k}, \label{X}
\end{align}
\end{subequations}
are the operators for the momentum and position, respectively, and $\nabla_{\mathbf k}$ is the gradient operator with respect to $\mathbf k$.
Clearly, if the constraint (\ref{TC}) were absent, the position (\ref{X}) would be commutative and the OAM (\ref{L}) could be canonically quantized.
In a previous paper \cite{Li17} I deduced from the constraint (\ref{TC}) a quantum gauge. It characterizes the entanglement of the polarization with the momentum in such a way that only in a particular gauge can the polarization be represented independently of the momentum by the Pauli matrices and therefore be canonically quantized. One of the observable effects of the gauge degree of freedom was also clarified.
In this paper I will go on to discuss how the quantum gauge determines the quantization of the OAM. Here are the main results.

The quantum gauge identified in Ref. \cite{Li17} turns out to be a Berry gauge. It determines a reference point relative to which the position is commutative and is canonically conjugate to the momentum. As a result, only when considered in a particular Berry gauge can the momentum be regarded as canonical.
The same as the polarization quantum number, the extrinsic quantum numbers that follow from the canonical position and momentum are physically meaningful only when they are associated with a particular Berry gauge. This gives rise to another observable effect of the Berry-gauge degree of freedom.
The OAM about that reference point is a canonical OAM satisfying the canonical commutation relation (\ref{CCR-L}). The so-called eigenvalue of the OAM about the origin of the laboratory reference system is actually the eigenvalue of its canonical part in the case in which the Berry-gauge degree of freedom of a paraxial beam is perpendicular to its propagation direction.

\section{Brief review of the quantum-gauge representation}

For later convenience, we briefly review how the quantum-gauge representation is deduced from the constraint (\ref{TC}).
As is known, the constraint (\ref{TC}) makes it possible to expand the vector wavefunction $\mathbf f$ at each value of $\mathbf k$ in terms of two orthogonal base vectors.
Let be $\mathbf u$ and $\mathbf v$ a pair of mutually perpendicular unit vectors that form with $\mathbf w$ a right-handed Cartesian system
$\mathbf{uvw}$, satisfying
\begin{equation}\label{triad}
    \mathbf{u} \times \mathbf{v} =\mathbf{w}, \quad
    \mathbf{v} \times \mathbf{w} =\mathbf{u}, \quad
    \mathbf{w} \times \mathbf{u} =\mathbf{v}.
\end{equation}
Such unit vectors are not unique. But they can be fixed by a constant unit vector $\mathbf I$ in the following way,
\begin{equation}\label{basis}
    \mathbf{u}=\mathbf{v} \times \frac{\mathbf k}{k},             \quad
    \mathbf{v}=\frac{\mathbf{I} \times\mathbf{k}}{|\mathbf{I} \times\mathbf{k}|}.
\end{equation}
Choosing $\mathbf u$ and $\mathbf v$ as the base vectors as usual, we may write $\mathbf f$ as
\begin{equation*}
    \mathbf{f} =f_1 \mathbf{u}
               +f_2 \mathbf{v}.
\end{equation*}
Putting the two expansion coefficients $f_1$ and $f_2$ together to form a two-component wavefunction
$
    \tilde{f}=\bigg(\begin{array}{c}
                      f_1 \\
                      f_2
                    \end{array}
              \bigg)
$,
we can rewrite $\mathbf f$ as
\begin{equation}\label{QUT-1}
    \mathbf{f} =\varpi \tilde{f},
\end{equation}
where
$
\varpi=(
         \begin{array}{cc}
           \mathbf{u} & \mathbf{v} \\
         \end{array}
       )
$
is a 3-by-2 matrix consisting of the two base vectors and vectors of three components such as $\mathbf u$ and $\mathbf v$ are expressed as column matrices.
Since the matrix $\varpi$ guarantees that the vector wavefunction (\ref{QUT-1}) obeys the constraint (\ref{TC}), the two-component wavefunction is no longer subject to such a condition.
It constitutes a gauge representation in the following sense.

The matrix $\varpi$ has property
\begin{equation}\label{unitarity-2}
    \varpi^{\dag} \varpi =I_2,
\end{equation}
where $I_2$ is the 2-by-2 unit matrix and the superscript $\dag$ stands for the conjugate transpose.
Multiplying both sides of Eq. (\ref{QUT-1}) by $\varpi^{\dag}$ from the left and using Eq. (\ref{unitarity-2}), we get
\begin{equation}\label{QUT-2}
    \tilde{f} =\varpi^\dag \mathbf{f}.
\end{equation}
Since the vector wavefunction is defined with respect to the laboratory reference system, it follows from this equation that the two-component wavefunction is defined with respect to the local reference system $\mathbf{uvw}$ that is determined by the unit vector $\mathbf I$ through Eqs. (\ref{basis}).
We shall see that such a property of the two-component wavefunction will have a decisive effect on the quantization of the OAM.
Once the unit vector $\mathbf I$ is chosen, the two-component wavefunction will have a one-to-one correspondence with the vector wavefunction via Eq. (\ref{QUT-2}), constituting a two-component representation.
But for a given vector wavefunction $\mathbf f$, the two-component wavefunction (\ref{QUT-2}) is dependent on the choice of the unit vector $\mathbf I$. This shows that the two-component representation is a gauge representation. The unit vector $\mathbf I$ is the degree of freedom to fix the gauge.
To see how the two-component wavefunction depends on the gauge, we choose a different gauge, $\mathbf{I}'$ say, to fix the transverse axes of the local reference system,
\begin{equation*}
    \mathbf{u}'=\mathbf{v}' \times \frac{\mathbf k}{k},   \quad
    \mathbf{v}'=\frac{\mathbf{I}' \times\mathbf{k}}
                     {|\mathbf{I}' \times \mathbf{k}|}.
\end{equation*}
In this case, the two-component wavefunction is given by
\begin{equation}\label{QUT-2'}
    \tilde{f}'=\varpi'^{\dag} \mathbf{f},
\end{equation}
where
$
\varpi'=(\begin{array}{cc}
            \mathbf{u}' & \mathbf{v}'
          \end{array}
        ).
$
It is remarked \cite{Mandel} that the new transverse axes
$\mathbf{u}'$ and $\mathbf{v}'$
are related to the old ones $\mathbf{u}$ and $\mathbf{v}$ by a rotation about $\mathbf k$.
Letting be $\Phi(\mathbf{k}; \mathbf{I}, \mathbf{I}')$ the $\mathbf k$-dependent rotation angle, such a rotation can be expressed compactly as
\begin{equation}\label{rotation-pi}
    \varpi' =\varpi \exp (-i \hat{\sigma}_3 \Phi),
\end{equation}
where
$
    \hat{\sigma}_3=\bigg(\begin{array}{cc}
                           0 & -i \\
                           i &  0
                         \end{array}
                   \bigg).
$
Substituting Eq. (\ref{rotation-pi}) into Eq. (\ref{QUT-2'}) and noticing Eq. (\ref{QUT-2}), we find
\begin{equation}\label{GT-f}
    \tilde{f}'=\exp (i \hat{\sigma}_3 \Phi) \tilde{f}.
\end{equation}
This is the gauge transformation of the two-component wavefunction. We will see in the following section that so identified quantum gauge is a Berry gauge.

\section{Canonical position and Berry gauge}\label{BGI}

\subsection{Canonical position in a gauge representation}

According to Eq. (\ref{QUT-2}), the operators for the momentum and position in the gauge representation are transformed from their counterparts (\ref{XandP}) in the vector representation as follows,
\begin{subequations}
\begin{align}
  \hat{\mathbf p}& =\varpi^{\dag} \hat{\mathbf P} \varpi=\hbar \mathbf{k}, \label{p} \\
  \hat{\mathbf x}& =\varpi^{\dag} \hat{\mathbf X} \varpi
                    =\hat{\boldsymbol \xi} +\hat{\mathbf b},               \label{x}
\end{align}
\end{subequations}
where
\begin{subequations}
\begin{align}
  \hat{\boldsymbol \xi} & =i \nabla_\mathbf{k},              \label{xi} \\
  \hat{\mathbf b}       & =i \varpi^{\dag} (\nabla_\mathbf{k} \varpi). \label{bI}
\end{align}
\end{subequations}
The momentum operator (\ref{p}) remains the same as in the vector representation. But the position operator (\ref{x}) splits into two parts.
The first part (\ref{xi}) has the same form as the position operator (\ref{X}) in the vector representation.
But because no conditions such as Eq. (\ref{TC}) exist for the two-component wavefunction, it is commutative,
\begin{equation}\label{CR-xi}
    [\hat{\xi}_i, \hat{\xi}_j] =0,
\end{equation}
the same as the momentum operator (\ref{p}),
\begin{equation}\label{CR-p}
    [\hat{p}_i, \hat{p}_j] =0.
\end{equation}
In addition, it has the following commutation relation with the momentum,
\begin{equation}\label{CR-xip}
    [\hat{\xi}_i,\hat{p}_j] =i \hbar \delta_{ij}.
\end{equation}
The second part (\ref{bI}) is solely determined by the matrix $\varpi$.  A straightforward calculation gives
\begin{equation}\label{b-sigma}
    \hat{\mathbf b}=\hat{\sigma}_3 \mathbf{A},
\end{equation}
where $\hat{\sigma}_3$ is the Pauli matrix in (\ref{rotation-pi}) and
\begin{equation}\label{AI}
    \mathbf{A}(\mathbf{k}, \mathbf{I})=\frac{\mathbf{I} \cdot \mathbf{k}}
              {k |\mathbf{I} \times \mathbf{k}|} \mathbf{v}
\end{equation}
is a vector function depending on the gauge degree of freedom $\mathbf I$.
Commuting with the Hamiltonian, this part represents a constant of motion. Moreover, its Cartesian components also commute,
\begin{equation}\label{CR-b}
    [\hat{b}_i, \hat{b}_j]=0.
\end{equation}

Eqs. (\ref{CR-xi})-(\ref{CR-xip}) are nothing but the canonical commutation relations between $\hat{\boldsymbol \xi}$ and $\hat{\mathbf p}$.
In other words, the operator $\hat{\boldsymbol \xi}$ represents such a position that is canonically conjugate to the momentum, called the canonical position.
The key point here is that the canonical position is not the laboratory position $\hat{\mathbf x}$. It is instead measured relative to the reference point $\hat{\mathbf b}$ that is determined by the gauge degree of freedom through the vector function $\mathbf A$. This is however in consistency with our previous observation: the two-component wavefunction is defined with respect to the local reference system $\mathbf{uvw}$ that is determined by the gauge degree of freedom.
Now that the reference point for the canonical position is determined by the gauge degree of freedom, the momentum can be regarded as canonical only when it is considered in a particular gauge.
As a result, the quantum numbers that follow from the commutation relations (\ref{CR-xi})-(\ref{CR-xip}) are physically meaningful only when they are associated with a particular gauge.
With the help of Eqs. (\ref{CR-xi}) and (\ref{CR-b}), it is not difficult to find
\begin{equation}\label{CR-x}
    [\hat{x}_i,\hat{x}_j]=i \hat{\sigma}_3 \epsilon_{ijk}B_k,
\end{equation}
where
\begin{equation}\label{B}
    \mathbf{B} =\nabla_\mathbf{k} \times \mathbf{A}
               =-\frac{\mathbf w}{k^2}, \quad \mathbf{w} \neq \pm \mathbf{I}.
\end{equation}
This shows that the constraint (\ref{TC}) on the vector wavefunction reflects the noncommutativity of the laboratory position.
To understand this commutation relation, let us first discuss the meaning of the quantity that is represented by $\hat{\sigma}_3$ by studying the spin operator in the gauge representation.

\subsection{The spin is independent of the gauge}

The spin operator in the vector representation takes the form \cite{Cohen, Akhiezer, Li09-1, Li16}
$
\hat{\mathbf S}= \hbar \hat{\mathbf \Sigma},
$
where
$(\hat{\Sigma}_k)_{ij} =-i \epsilon_{ijk}$
with $\epsilon_{ijk}$ the Levi-Civit\'{a} pseudotensor.
According to Eq. (\ref{QUT-2}), the spin operator in the gauge representation is given by
\begin{equation*}
    \hat{\mathbf s} =\hbar \varpi^{\dag} \hat{\mathbf \Sigma} \varpi.
\end{equation*}
Upon decomposing the vector operator $\hat{\boldsymbol \Sigma}$ in the local reference system $\mathbf{uvw}$ as
\begin{equation*}
    \hat{\boldsymbol \Sigma}
       =(\hat{\boldsymbol \Sigma} \cdot \mathbf{u}) \mathbf{u}
       +(\hat{\boldsymbol \Sigma} \cdot \mathbf{v}) \mathbf{v}
       +(\hat{\boldsymbol \Sigma} \cdot \mathbf{w}) \mathbf{w}
\end{equation*}
and taking Eqs. (\ref{triad}) into account, we find
\begin{equation}\label{s-sigma}
    \hat{\mathbf s}= \hbar \hat{\sigma}_3 \mathbf{w},
\end{equation}
where
$
\hat{\sigma}_3=\varpi^{\dag} (\hat{\mathbf \Sigma} \cdot \mathbf{w}) \varpi.
$
Here we arrive in a simple way at the well-known conclusion that the spin of the photon lies entirely on its propagation direction. The Pauli matrix $\hat{\sigma}_3$ represents essentially the magnitude of the spin, known as the helicity.
As a result, the Cartesian components of the spin commute,
\begin{equation*}
    [\hat{s}_i, \hat{s}_j]=0.
\end{equation*}
This is what van Enk and Nienhuis obtained in a second-quantization framework ~\cite{EN, Enk}.

A comparison between the spin operator (\ref{s-sigma}) and the polarization operator \cite{Li17} in the gauge representation gives a clear answer to the long-standing question \cite{Darw, Berry98, Mili} about the difference between the spin and the polarization.
In the first place, the spin is only the longitudinal part of the polarization. In the second place, because the helicity is the generator of the gauge transformation (\ref{GT-f}), the spin is independent of the choice of the gauge. This is in contrast with the polarization that is dependent on the choice of the gauge.

\subsection{The quantum gauge is a Berry gauge}

Now we are in a position to show that the quantum gauge is a Berry gauge. Substituting Eq. (\ref{b-sigma}) into Eq. (\ref{x}), we have
\begin{equation}\label{x-A}
    \hat{\mathbf x}=\hat{\boldsymbol \xi} +\hat{\sigma}_3 \mathbf{A}.
\end{equation}
If the laboratory position $\hat{\mathbf x}$ is regarded as the analog of the kinetic momentum of an electrically charged particle in a magnetic field and the canonical position $\hat{\boldsymbol \xi}$ is regarded as the analog of the canonical momentum of the particle~\cite{Barut}, then the helicity $\hat{\sigma}_3$ can be regarded as the analog of the electric charge of the particle and the vector function $\mathbf{A}$ can be regarded as the analog of the vector potential of the magnetic field.
That is, $\mathbf{A}$ appears as the gauge potential of the gauge field (\ref{B}).
Indeed, the commutation relation (\ref{CR-x}) of the laboratory position that depends on the gauge field (\ref{B}) is analogous to the commutation relation of the kinetic momentum of the charged particle that depends on the magnetic field.

Furthermore, consider a different gauge representation $\mathbf{I}'$. The operator in this case for the laboratory position becomes
\begin{equation*}
    \hat{\mathbf x}' =\varpi'^{\dag} \hat{\mathbf X} \varpi'
                     =\hat{\boldsymbol \xi}+\hat{\mathbf b}',
\end{equation*}
where
\begin{equation}\label{bI'}
    \hat{\mathbf b}'=\hat{\sigma}_3 \mathbf{A}'
\end{equation}
and $\mathbf{A}'=\mathbf{A}(\mathbf{k}, \mathbf{I}')$.
In addition, from Eq. (\ref{bI}) it follows that
$\hat{\mathbf b}'=i \varpi'^\dag(\nabla_\mathbf{k} \varpi')$.
With the help of Eqs. (\ref{rotation-pi}) and (\ref{unitarity-2}), we get
\begin{equation}\label{b'-b}
    \hat{\mathbf b}'
   =\hat{\mathbf b}+\hat{\sigma}_3 \nabla_{\mathbf k} \Phi.
\end{equation}
A comparison of Eqs. (\ref{bI'}) and (\ref{b-sigma}) gives
\begin{equation}\label{GT-A}
    \mathbf{A}' =\mathbf{A}+\nabla_{\mathbf k} \Phi,
\end{equation}
which is a gauge transformation on the gauge potential with $\Phi$ the corresponding gauge function. This shows that the unit vector $\mathbf I$ is indeed a gauge degree of freedom. The meaning of it is explained as follows.

The gauge field (\ref{B}) corresponding to the gauge potential (\ref{AI}) is a Berry-gauge field \cite{Berry84} that is produced by a magnetic monopole \cite{Dirac31} of unit strength in momentum space \cite{Bia-B1987, Fang}.
The gauge degree of freedom $\mathbf I$, called the Berry-gauge degree of freedom, indicates the orientation of the monopole's Dirac string.
The gauge transformation (\ref{GT-A}) serves to reorient the Dirac string from $\mathbf I$ to $\mathbf{I}'$ and is thus a singular transformation \cite{Milton}.
In a word, each unit vector $\mathbf I$ represents a particular Berry gauge through the Berry-gauge potential (\ref{AI}).
This result leads us to an unusual conclusion that the commutation relation (\ref{CR-x}) of the laboratory position is dependent on the Berry gauge, implying an observable quantum effect of the Berry-gauge degree of freedom in this aspect.

\subsection{Dependence of the barycenter on the Berry gauge}

To show this, we consider in the Berry-gauge representation the simultaneous eigenstate of the helicity and momentum the wavefunction of which is given by
\begin{equation}\label{EF}
    \tilde{f}_{\sigma, \mathbf{k}_0}
   =\tilde{\alpha}_\sigma \delta^3 (\mathbf{k}-\mathbf{k}_0) \exp(-i \omega t),
\end{equation}
where
$
\tilde{\alpha}_\sigma =\frac{1}{\sqrt{2}} \bigg(
                                                \begin{array}{c}
                                                      1 \\
                                                      i \sigma \\
                                                \end{array}
                                          \bigg)
$
is the eigenfunction of the helicity operator $\hat{\sigma}_3$ with eigenvalue $\sigma =\pm 1$ and $\mathbf{k}_0$ is the eigenvalue of the momentum.
Apparently, it is the eigenstate of the operator (\ref{b-sigma}) with the eigenvalue,
\begin{equation}\label{b-EV}
    \mathbf{b}_{\sigma, \mathbf{k}_0; \mathbf{I}}
   =\sigma \frac{\mathbf{I} \cdot \mathbf{k}_0}
                {k_0 |\mathbf{I} \times \mathbf{k}_0|^2}
                \mathbf{I} \times \mathbf{k}_0,
\end{equation}
where $k_0=|\mathbf{k}_0|$.
Observing that the expectation value of the canonical position in this state vanishes,
$\langle \hat{\boldsymbol \xi} \rangle =0$,
the eigenvalue (\ref{b-EV}) is nothing but the barycenter of the photon in the sense that it is equal to the expectation value of the laboratory position,
$
\langle \hat{\mathbf x} \rangle=\mathbf{b}_{\sigma, \mathbf{k}_0; \mathbf{I}}.
$
It illustrates that the same set of quantum numbers $\{ \sigma, \mathbf{k}_0 \}$ in different Berry gauge stands for different eigenstate, having different barycenter.
What deserves emphasizing is that the helicity has crucial contribution to the barycenter.
This is why the so-called spin Hall effect of light \cite{Onoda, Hosten} can be quantitatively explained by the change of the Berry-gauge degree of freedom \cite{Li09-2}.

\section{Interpretation of the eigenvalue of the OAM}\label{OAM}

Now we are ready to deal with the OAM.
The same as the position operator (\ref{x}), the operator in the Berry-gauge representation for the OAM about the origin of the laboratory reference system also splits into two parts,
\begin{equation}\label{l}
    \hat{\mathbf l}= \varpi^\dag \hat{\mathbf L} \varpi
                   = \hat{\boldsymbol \lambda} +\hat{\mathbf m}.
\end{equation}
The first part
\begin{equation}\label{lambda}
    \hat{\boldsymbol \lambda}=-\hat{\mathbf p} \times \hat{\boldsymbol \xi}
\end{equation}
represents the OAM of the photon about the reference point $\hat{\mathbf b}$, the barycenter. Thanks to the canonical commutation relations (\ref{CR-xi})-(\ref{CR-xip}), it satisfies the canonical commutation relation of the angular momentum,
\begin{equation}\label{CR-lambda}
    [\hat{\lambda}_i, \hat{\lambda}_j] =i \hbar \epsilon_{ijk} \hat{\lambda}_k,
\end{equation}
and will be referred to as the canonical OAM. Clearly, it is a constant of motion,
\begin{equation}\label{CR-lambda and omega}
    [\hat{\boldsymbol \lambda}, \omega] =0.
\end{equation}
The second part
\begin{equation}\label{mI}
    \hat{\mathbf m}
   \equiv \hat{\mathbf b} \times \hat{\mathbf p}
   =\hbar \hat{\sigma}_3 \frac{\mathbf{I} \cdot \mathbf{k}}{|\mathbf{I} \times \mathbf{k}|} \mathbf{u}
\end{equation}
represents the OAM of the photon concentrated at the barycenter.
Different from the first part, its Cartesian components commute,
\begin{equation}\label{CR-m}
    [\hat{m}_i, \hat{m}_j]=0.
\end{equation}
But the same as the first part, it is a constant of motion,
\begin{equation}\label{CR-m and omega}
    [\hat{\mathbf m}, \omega] =0.
\end{equation}
As a result of Eqs. (\ref{CR-lambda and omega}) and (\ref{CR-m and omega}), the total OAM is a constant of motion, too.

It is interesting to note that the expression (\ref{l}) for the total OAM can find its counterpart in classical mechanics \cite{Goldstein}: the angular momentum of a system about a reference point is the angular momentum of the system concentrated at the barycenter plus the angular momentum of the system about the barycenter.
With the help of Eqs. (\ref{CR-lambda}) and (\ref{CR-m}), it is not difficult to find for the commutation relation of the total OAM in any Berry gauge,
\begin{equation*}
    [\hat{l}_i, \hat{l}_j]
   =i \hbar \epsilon_{ijk} (\hat{l}_k-\hat{s}_k),
\end{equation*}
in agreement with what van Enk and Nienhuis obtained in a second-quantization framework \cite{EN, Enk}.
This shows that the total OAM cannot be canonically quantized. Only its first part, the canonical OAM, can.
In addition, the dependence of the second part on the helicity explains why the total angular momentum of a non-paraxial beam cannot be separated into helicity-independent OAM and helicity-dependent spin \cite{Barnett-A}.

A paraxial light beam is a photon state in which all the plane-wave component travels in almost the same direction, the propagation axis. If the Berry-gauge degree of freedom of a paraxial beam is perpendicular to the propagation axis, it follows from Eq. (\ref{mI}) that
$\hat{\mathbf m} \approx 0$.
In this case, the total OAM is approximately the canonical OAM,
$\hat{\mathbf l} \approx \hat{\boldsymbol \lambda}$.
The paraxial beam that was discussed by Allen et. al. \cite{Allen} is just such a beam \cite{Li09-1}. The so-called eigenvalue of the OAM is actually the eigenvalue of the canonical OAM.

\section{Conclusions and comments}\label{remarks}

In conclusion, the quantum gauge that is deduced from the transversality condition (\ref{TC}) is shown to be a Berry gauge. It determines the reference point (\ref{b-sigma}) for the canonical position. This explains why the OAM about the origin of the laboratory reference system cannot be canonically quantized. Meanwhile, the difference between the spin and the polarization is also discussed.
The same as the polarization quantum number \cite{Li17}, the quantum numbers that follow from the canonical commutation relations (\ref{CR-xi})-(\ref{CR-xip}) are physically meaningful only when they are associated with a particular Berry gauge. It is thus inferred that only in a Berry gauge can the photon field be canonically quantized.
More importantly, the same eigenfunction in different Berry-gauge representation describes different photon state. Specifically, the simultaneous eigenstate (\ref{EF}) of the helicity and momentum in different Berry gauge has different barycenter, a phenomenon that was already observed in the spin Hall effect of light.
It is expected that the findings reported here will open up a new area of research on the photon angular momentum as well as on the quantization of the photon field.

\section*{Acknowledgments}

The author is indebted to Vladimir Fedoseyev and Zihua Xin for their helpful discussions.

\end{document}